\documentclass[aps,reprint,twocolumn,superscriptaddress,nofootinbib]{revtex4-1}

\pdfoutput=1

\usepackage{tabularx}

\newcolumntype{Y}{>{\centering\arraybackslash}X}

\usepackage{slashed}

\usepackage[normalem]{ulem}

\usepackage{color}

\usepackage{epsfig}
\usepackage{epstopdf}
\usepackage{epsf}
\input{epsf.sty}
\usepackage{graphicx,amsmath,amssymb}
\usepackage{epsfig}
\allowdisplaybreaks

\def\ei{\end{itemize}}
\def\be{\begin{equation}}
\def\ee{\end{equation}}
\newcommand{\bea}{\begin{eqnarray}}
\newcommand{\eea}{\end{eqnarray}}

\usepackage{bbm,bm,amsmath,amssymb}

\usepackage{hyperref}

\begin{document}

\title{Dark Disk Substructure and Superfluid Dark Matter}

\author{Stephon Alexander}
\email{stephon\_alexander@brown.edu}
\affiliation{Department of Physics, Brown University, Providence, RI, USA. 02912}
\author{Jason J. Bramburger}
\email{jason\_bramburger@brown.edu}
\affiliation{Division of Applied Mathematics, Brown University, Providence, RI, USA. 02906}
\author{Evan McDonough}
\email{evan\_mcdonough@brown.edu}
\affiliation{Department of Physics, Brown University, Providence, RI, USA. 02912}

 \begin{abstract}
Dark matter substructure has the potential to discriminate between broad classes of dark matter models. With this in mind, 
 we construct novel solutions to the equations of motion governing condensate dark matter candidates, namely axion Bose-Einstein condensates and superfluids. These solutions are highly compressed along one axis and thus have a disk-like geometry. We discuss linear stability of these solutions, consider the astrophysical implications as a large-scale dark disk or as small scale substructure, and find a characteristic signal in strong gravitational lensing. This adds to the growing body of work that indicates that the morphology of dark matter substructure is a powerful probe of the nature of dark matter.
\end{abstract}

\maketitle

\section{Introduction}

Strong lensing observations, in agreement with expectations from hierarchical stucture formation, have revealed the existence of dark matter \emph{substructure} \cite{Hezaveh:2016ltk}, that is, gravitationally bound clumps distinct from the halo. More recently, observations of the distribution of accreted stars in the local halo (SDSS-Gaia DR2 \cite{Brown:2018dum}) indicate the presence of substructure  \cite{Bonaca:2018fek,Price_Whelan_2018}, building on earlier revelations that a non-trivial fraction of dark matter in the local dark matter halo is in kinematic substructure \cite{Necib:2018iwb,Myeong:2017skt,Evans:2018bqy}.

Meanwhile, the nature of dark matter remains elusive, and the strongest evidence for dark matter remains gravitational.  Indeed, there is mounting evidence that dark matter substructure is a powerful tool to discriminate between classes of dark matter models, e.g. an intriguing feature of so-called `Fuzzy Dark Matter' \cite{Hu:2000ke,Hui:2016ltb} is a resolution of the small-scale crises of $\Lambda$CDM \cite{Bullock:2017xww}, namely the core-cusp \cite{Amorisco:2011hb,Amorisco:2012rd} and missing satellites  \cite{Bullock:2017xww}\footnote{Though recent evidence indicates that there is no such problem in need of solving \cite{PhysRevLett.121.211302}.} problems.

Particle dark matter predicts spherical sub-halos on all scales \cite{Kauffmann:1993gv}, while \cite{Fan:2013tia, Fan:2013yva} has argued that a strongly interacting subcomponent can lead to a ``dark disk'' aligned with the visible disk\footnote{In the same context \cite{Chang:2018bgx} recently argued for the existence of spherical compact objects.}.  Condensate dark matter scenarios, such as axion Bose-Einstein condensates \cite{Hu:2000ke,Sikivie:2009qn,Hui:2016ltb}, bosonic superfluids \cite{Berezhiani:2015bqa,Ferreira:2018wup,Alexander:2018lno} and fermionic superfluids \cite{Alexander:2016glq,Alexander:2018fjp}, also exhibit substructure, and to-date the known forms are spherical clumps \cite{2011PhRvD84d3531C,2011PhRvD84d3532C,Hui:2016ltb,Croon:2018ybs,Schiappacasse:2017ham} or axion miniclusters \cite{Hogan:1988mp},  and superfluid vortices \cite{RindlerDaller:2011kx}. All three condensate systems, in the non-relativistic limit, are described by the same set of equations, exact solutions to which should exist as dark matter substructure.  The morphology of this substructure, insofar as it differs from that of particle dark matter, in principle provides a signature of condensate dark matter scenarios. 

However, it is a notoriously difficult task to find bound state solutions without spherical symmetry, see e.g.~footnote 21 of \cite{Hui:2016ltb}. With this in mind, in this letter we study a novel form of substructure that can exist in condensate models. We look for and find \emph{disc-like} structures, that can exist both as a large-scale dark disk or else as isolated substructures. This investigation reveals new solutions to old equations, and adds qualitatively new results to the existing mathematical literature.

 \section{The Equations of Motion of Non-Relativistic Dark Matter}
\label{sec:sols}

The coupled Gross-Pitaevskii (non-linear Schr\"{o}dinger) and Poisson equations emerge in the non-relativistic limit of a classical scalar field theory interacting with gravity. This arises in an astrophysical context as cold dark matter at high number density, e.g.~an ultralight scalar \cite{Hu:2000ke,Hui:2016ltb}. The non-relativistic limit is defined via the decomposition
\be
\phi(x,t) = \sqrt{\frac{\hbar^3 c}{2m}} \left( \psi(x,t) e^{-i m c^2 t/\hbar} + c.c.\right) ,
\ee
and the limit $|\ddot{\psi}| \ll m c^2 |\dot{\psi}|/\hbar$ \cite{Hui:2016ltb}. The equations of motion are then given by\footnote{It is is also possible to consider more exotic superfluids, e.g.~\cite{Berezhiani:2015bqa}, but here we will focus on the simplest case.}, 
\bea
i\dot{\psi} &&= \left(-\dfrac{1}{2m}\nabla^{2}+mV- \dfrac{\lambda}{m^{2}} \vert\psi\vert^{2}\right)\psi , \\
\nabla^{2}V &&=4 \pi G m \vert\psi\vert^{2}, 
\eea 
where $\nabla^2$ is the Laplacian in three spatial dimensions, and we set $c,\hbar=1$. Here $\lambda > 0$ corresponds to an attractive interaction. We will take $\lambda>0$ through-out. 

We will consider a static solution with an energy $E$, such that $\psi(x,t)= \tilde{\psi}(x) e^{- i E t}$. The equations of motion can be put in a dimensionless form via the redefinitions, $\vec{x} \equiv R \vec{\hat{x}}$ , $V - E \equiv V_0\hat{V}$ , $\tilde{\psi} \equiv \psi_0 \hat{\psi}$, with 
\be
\label{eq:rescalings}
R \equiv \frac{\sqrt{2} m_{pl} \sqrt{\lambda}}{m^2}, \quad \psi_0 \equiv \frac{m^{5/2}}{2 m_{pl} \lambda}, \quad V_0 \equiv \frac{m R^2 \psi_0 ^2}{2 m_{pl} ^2},
\ee 
where $m_{pl}$ is the reduced Planck mass. The characteristic distance scale $R$ is thus set by the coupling $\lambda$ and the mass $m$.  The equations of motion are then given in terms of the hatted quantities by,
\begin{equation} \label{eq:NormalizedEq}
	\begin{split}
		\left(-\hat{\nabla}^{2}+\hat{V} - |\hat{\psi}|^{2} \right)\hat{\psi} &= 0 ,\\
		\hat{\nabla}^{2} \hat{V} &= |\hat{\psi}|^{2} ,
	\end{split}
\end{equation}
where $\hat{\nabla}$ denotes the Laplacian with respect to the hatted coordinates, $\nabla^2 \equiv \frac{\partial^2 }{\partial \hat{x}^2} + \frac{\partial^2 }{\partial \hat{y}^2} + \frac{\partial^2 }{\partial \hat{z}^2} $.

We note that by absorbing $\lambda$ into the rescalings \eqref{eq:rescalings}, the bound state solutions found here are intrinsically self-interacting. Axion dark matter candidates generically have extremely small self-interactions, for example, QCD axion dark matter with mass $m_a \simeq 10^{-6} {\rm eV}$ and decay constant $f_a \simeq 10^{12} {\rm GeV}$  \cite{Preskill:1982cy} gives $\lambda  \simeq m_a^2 /f_a^2 = 10^{-54}$. In the extreme case of Fuzzy Dark Matter, the self-interaction can be as small as $\lambda=10^{-96}$ \cite{Desjacques:2017fmf}. Similarly, in this work we will consider $\lambda \ll 1$.

\section{Solutions Away from Spherical Symmetry}

It is conventional to solve the above equations assuming spherical symmetry of the solution \cite{Croon:2018ybs,Schiappacasse:2017ham,2011PhRvD84d3531C,2011PhRvD84d3532C}\footnote{See also \cite{1985MNRAS} for a related analysis of a relativistic scalar field.}; see also \cite{Moroz:1998dh,Tod,Tod2,Harrison:2003,Choquard:2008} for the case $\lambda=0$. In our work, we are interested in disk-like solutions, which instead have an \emph{axial} symmetry, i.e.~invariance with respect to rotations about a fixed axis. As previously noted, it is a difficult to task to find bound state solutions to \eqref{eq:NormalizedEq} without spherical symmetry. This difficulty is due in large part to the non-linearity of \eqref{eq:NormalizedEq} and to the fact that axially symmetric bound state solutions must have (at least) two independent variables. However, symmetry-reduced phenomena are known to exist in closely related systems (e.g.~\cite{2006PhRvL}), and thus it is not unreasonable to propose that such stable or metastable solutions exist in three dimensions with gravity.

To this end, we note that a simple class of disk-like geometries in $\mathbb{R}^3$ can be described in terms of a squeezed radial coordinate
\be
\hat{r}_{\rm sq} ^2 = \hat{x}^2 + \hat{y}^2 + (D_{\rm sq}-1) \hat{z}^2 ,
\ee
with real-valued $D_{\rm sq}>2$.  For large $D_{\rm sq}$, functions of the form $\psi(x,y,z) = \psi(\hat{r}_{\rm sq})$ which decay at infinity are disk-like in $\mathbb{R}^3$. Interestingly, the Laplacian acting on a wavefunction $\psi(\hat{r}_{\rm sq})$ takes a simple form when expanded in powers of $\hat{z}/D_{\rm sq}$:
 \begin{equation} \label{eq:Laplz0}
	\nabla^2 \equiv \frac{\partial^2 }{\partial \hat{r}^2 _{\rm sq}} + \frac{D_{\rm sq}}{\hat{r}_{\rm sq}}  \frac{\partial}{\partial \hat{r}_{\rm sq}} + \mathcal{O}\bigg(\frac{\hat{z}^2}{D^2 _{\rm sq}}\bigg), \quad \hat{z} \ll D_{\rm sq} ,
\end{equation}
When $D_{\rm sq}$ is a positive integer, one recognizes the leading order term of (\ref{eq:Laplz0}) as the radially symmetric Laplacian in $D_{\rm sq}+1$ spatial dimensions. Our present work allows for $D_{\rm sq}$ to be an arbitrary positive real number so that it may be interpreted as interpolating between dimensions corresponding to integer values of $D_{\rm sq}$. 

It follows that disk-like solutions may be described in the region $\hat{z}/D_{\rm sq} \ll 1$ as solutions to the radially symmetric equation (\ref{eq:Laplz0}) in very large spatial dimensions. Inserting \eqref{eq:Laplz0} into (\ref{eq:NormalizedEq}), and truncating at lowest order in $\hat{z}/D_{\rm sq}$, we have
\begin{equation}\label{eq:Final}
	\begin{split}
		-\bigg(\frac{\partial^2\psi }{\partial r^2} + \frac{D}{r}\frac{\partial \psi}{\partial r}\bigg)+V\psi - \psi^3 &=0,\\
	\frac{\partial^2 V}{\partial r^2} + \frac{D}{r}\frac{\partial V}{\partial r} &= \psi^2, 
	\end{split}
\end{equation} 
upon dropping the hats, ${\rm sq}$ subscripts, and restricting to real functions $\psi(r)$. Equation (\ref{eq:Final}) resembles a core differential equation obtained through the traditional spatial dynamics method of far-field/core decomposition (see e.g.~\cite{Scheel}). In the context of physics, this expansion is similar to conventional methods of electrodynamics, e.g.~in Fresnel Diffraction \cite{jackson}, and in a more modern context, a similar expansion appears in the famed Anti-de Sitter/Conformal Field Theory correspondence \cite{Maldacena1999}, where an Anti-de Sitter space emerges in the near-horizon limit of stack of coincident three-branes. We expect that solutions $\psi(r)$ of (\ref{eq:Final}) which rapidly decay to $0$, i.e.~are localized within the region $z\ll D$, well approximate solutions to the full equation \eqref{eq:NormalizedEq}. In what follows we will numerically solve \eqref{eq:Final}. 

\vspace{-.5cm}

\section{disk Solutions} \label{sec:Sols}

\vspace{-.3cm}


 We use the numerical fixed point algorithm described in the appendix to find solutions of (\ref{eq:Final}). Throughout this work we fix $V(0) = -1$, which corresponds physically to a choice of scalings \eqref{eq:rescalings}, but remark that solutions appear to exist for any choice of $V(0) < 0$, and therefore we leave an exploration on the dependence of the choice of $V(0)$ to a subsequent investigation. 

\begin{figure}[h!]
\includegraphics[width=0.45 \textwidth]{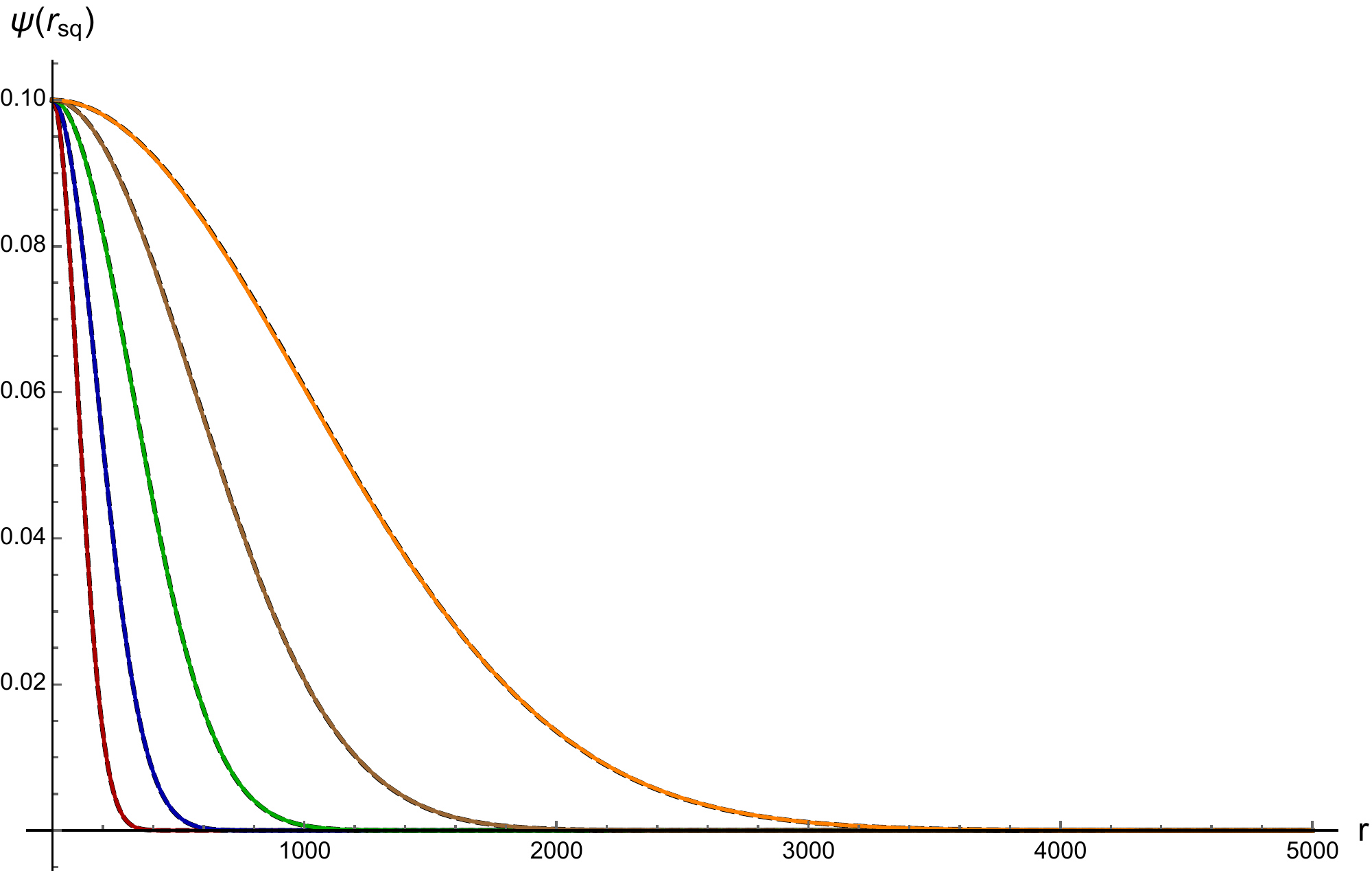}
\caption{A comparison between solutions $\psi(r)$ of (\ref{eq:Final}) and the fit (\ref{eq:Fit}) with $\psi(0) = 10^{-1}$ for various values of $D$. Numerical solutions are solid while the analytic fits are dashed, and from the left to right we have $D_{\rm sq}=10^4, 10^{4.5}, 10^{5}, 10^{5.5}, 10^6$. The axes are related to physical, dimensional, quantities through the rescalings \eqref{eq:rescalings}. }
\label{plot1}
\end{figure}

Our results indicate the existence of solutions of (\ref{eq:Final}) for arbitrarily large $D_{\rm sq} > 0$ for which $\hat{\psi}(\hat{r}_{\rm sq})$ monotonically decreases and approaches $0$ as $\hat{r}_{\rm sq} \to \infty$. Moreover, we have identified for $D_{\rm sq}$ a family of solutions which can be parametrized by their value at $r = 0$, simply written $\hat{\psi}(0)$, that are well fit by a Gaussian
\begin{equation}\label{eq:Fit}
	\hat{\psi}(\hat{r}_{\rm sq}) = \hat{\psi}(0)\mathrm{e}^{-\frac{\hat{r}^2 _{\rm sq }}{2D_{\rm sq}}},
\end{equation}
for every $\hat{\psi}(0) \in [0,1]$.  The fit to the wavefunction \eqref{eq:Fit}can be observed in Figure~\ref{plot1}, where we fix the prefactor $\hat{\psi}(0)$ to $10^{-1}$ and plot the wavefunction for increasing values of $D_{\rm sq}$ up from $10^4$.

\begin{figure}[h!]
\includegraphics[width=0.45 \textwidth]{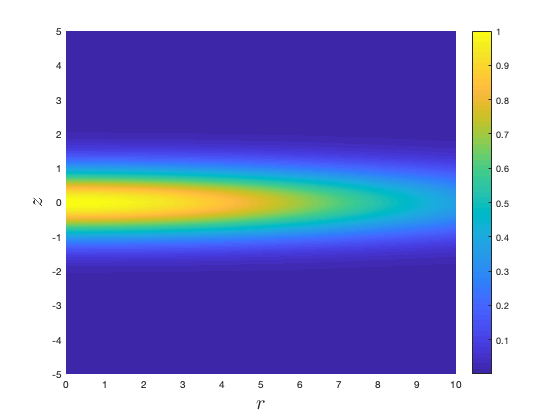}
\caption{The density profile \eqref{eq:Fitrho} in the ${r,z}$ plane for $D_{\rm sq}=100$, with the density normalized to $1$ at the origin, and with $R$ normalized to $1$. Here the coordinate $r$ is here defined as $r^2=x^2 + y^2$.}
\label{plotrho}
\end{figure}

This corresponds to a density profile in physical coordinates,
\be
\label{eq:Fitrho}
\rho(r) \equiv m |\psi|^2 =  \frac{m^6}{4 m_{pl}^2 \lambda^2} \hat{\psi}(0)^2 e^{ - \frac{x^2 + y^2}{ R^2 D_{\rm sq}}} e^{- \frac{z^2}{ R^2}},
\ee 
with $R$ given by \eqref{eq:rescalings}, and we have approximated $D_{{\rm sq}}/ (D_{\rm sq} -1)=1$ for notational simplicity in the exponents. The disk is thus characterized by a Gaussian density profile and a central density $\rho_0 \propto m^6/(m_{pl}^2 \lambda^2)$.  For large $D_{\rm sq}$, this is an extremely thin disk. This can be observed in Figure \ref{plotrho}, which plots the density \eqref{eq:Fitrho} in the $\{r\equiv \sqrt{x^2 + y^2},z\}$ plane for $D_{\rm sq}=100$, normalized to $1$ at the origin, and with $R$ normalized to $1$. This is considerably thinner then traditional disks found in astronomy and astrophysics, e.g. the thin-disc of the milky way has a thickness $300$ pc and radial extent $3$ kpc, for an overall compression factor $10$.

We note that the emergence of a family of solutions is not entirely surprising since in the absence of the self-interaction term $\hat{\psi}^3$ one may show that solutions exist for any value of $\hat{\psi}(0) > 0$ \cite{Tod2}. Hence, when $\hat{\psi}(0)$ is taken sufficiently small the self-interaction $\hat{\psi}^3$ is sub-dominant to the other terms in the Schr\"odinger equation, and therefor acts as a slight perturbation. Moreover, based on the approximate solution \eqref{eq:Fit}, it appears that finite-energy spherically symmetric solutions exist in an arbitrarily large number of spatial dimensions.

These solutions are characterized by a distance scale in physical, unhatted, coordinates,
\be
\label{Rc}
R_{c} \equiv \sqrt{2 D_{\rm sq}} R \simeq  32  \sqrt{D_{\rm sq} \lambda}  \left( \frac{\rm eV}{m}\right)^2 {\rm kpc} ,
\ee
which defines the `core radius', or planar extent, of the disk solutions. This can take a broad range of values, for example, for $D_{\rm sq}=10^5$, $m=10^{-10}$ eV, and $\lambda=10^{-45}$, this evaluates to 32 kpc. The disk \emph{thickness}, i.e.~the extent in the $z$-direction, is independent of $D_{\rm sq}$, and is therefore suppressed relative to the core radius \eqref{Rc} by a factor of $1/\sqrt{D_{\rm sq}}$. Thus, as anticipated, for large $D_{\rm sq}$ we find thin disk solutions. Moreover, the super-exponential decay in $\hat{z}$ is consistent with the assumption of localization of $\psi$ to the region $\hat{z}/D_{\rm sq} \ll 1$, providing an \emph{a posteori} justification of the use of the Laplacian \eqref{eq:Laplz0}.

The total mass of these disk solutions is given by,
\be
\label{MDD}
M_{DD} \equiv  \int \rho(x,y,z)\, d^3x \simeq 3 \times 10^{19} {\rm GeV} \hat{\psi}(0)^2 \frac{D_{\rm sq}}{\sqrt{\lambda}},
\ee
where again $x$ is the physical spatial coordinate, and we have used the analytic fit  \eqref{eq:Fitrho}. Most strikingly, the bound $\hat{\psi}(0)\leq 1$ provides an upper bound on the mass (\ref{MDD}) for each fixed $D_{\rm sq}$. Therefore, for a fixed $M_{DD} > 0$, we are able to obtain a lower bound on the squeezing factor $D_{\rm sq}$, which in terms of the conserved particle number $N\equiv M_{DD}/m$, is simply $D_{\rm sq} > 2 \pi^{3/2} N \sqrt{\lambda}  (m/m_{pl}) $.

The existence of this lower bound hints at the stability of these solutions. Namely, given a perturbation which leaves unchanged the boundary condition $\hat{\psi}(0)$, the conservation of $N$ forbids $D_{\rm sq}$ from dynamically relaxing to a small value, and therefore prevents a disk from relaxing to a spherical solution, the latter of which has $D=2$ in our convention. 

 This can be further probed by extremizing the total energy $H$, with respect to the particle number $N$ \cite{Croon:2018ybs,Schiappacasse:2017ham}. At fixed $\psi(0)$ this leads to an expression for the various contributions to $H$ in terms of $N$ and $D _{\rm sq}$, which exhibits a stable minimum at particle number $N \propto D _{\rm sq}$, as expected. 

\begin{figure}[h!]
\begin{center}
\includegraphics[width=0.45 \textwidth]{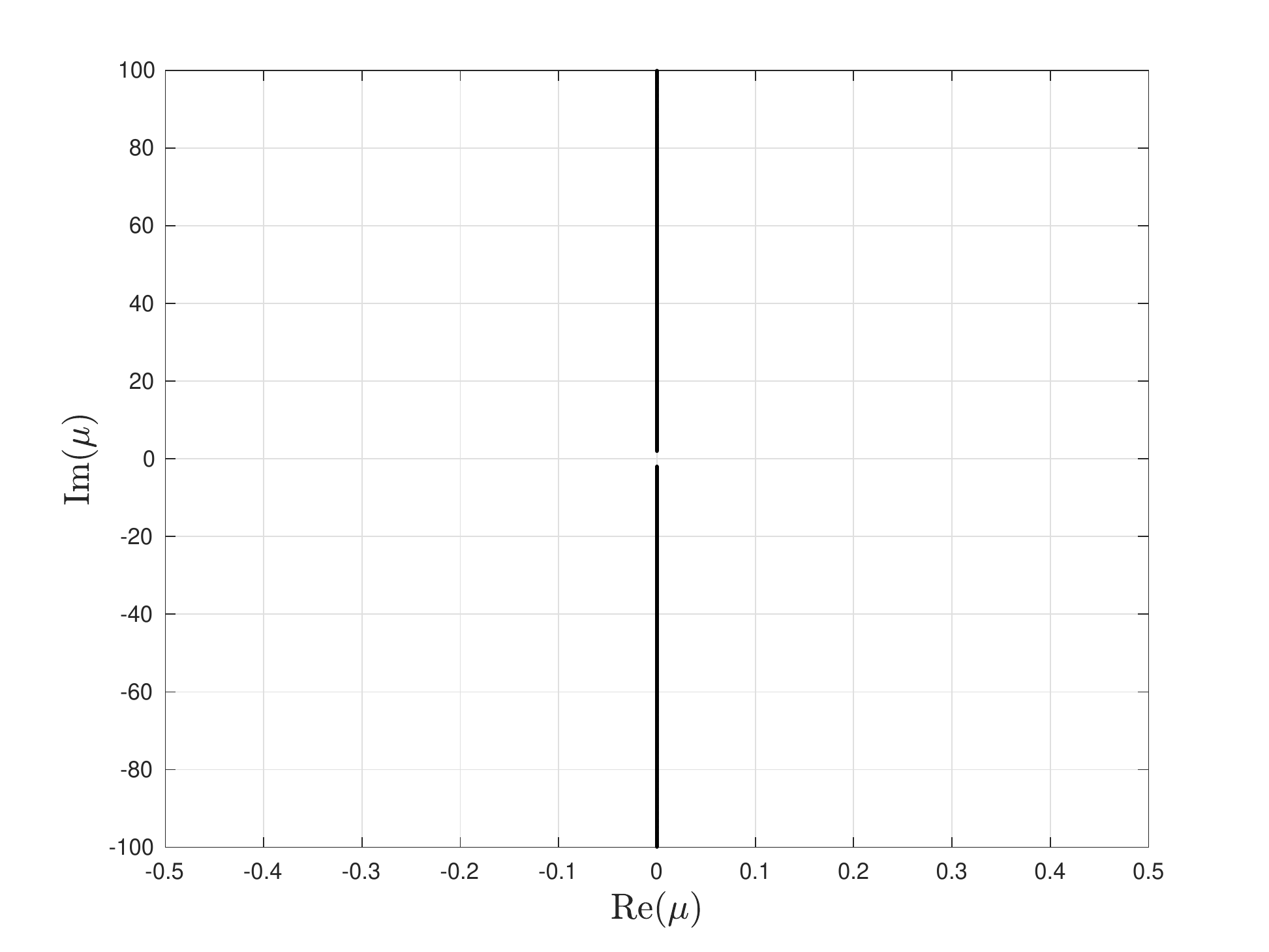}
\end{center}
\vspace{-.5cm}
\caption{Eigenvalues of \eqref{EigNumerics} for $D=10^3$ obtained using the methods outlined in the appendix. All eigenvalues lie on the imaginary axis indicating linear stability.}
\label{plot2}
\end{figure}

Finally, we turn to a numerical investigation of the linear stability of our solutions. In this preliminary investigation we only examine stability with respect to perturbations that are also axially symmetric, i.e.~are functions of just $r$ as well. The linear stability is carried out numerically using the methods outlined in the appendix. We find that these solutions have spectrum entirely contained on the imaginary axis, indicating linear stability. A plot of the spectrum for a solution with $D _{\rm sq}= 10^3$ is provided in Figure~\ref{plot2} and we remark that for all tested values of $D _{\rm sq}$ we have obtained linear stability as well. Of course this work only provides linear stability with respect to a limited class of perturbations, but does provide valuable information pertaining to these disk-like structures. We leave a full stability analysis to a follow-up analysis, and note that absolute stability is not required for such objects to be present in the galaxy, which is out of equilibrium \cite{Necib:2018iwb,Myeong:2017skt}.

\section{Astrophysical Implications}
\label{sec:astro}

 We have found self-gravitating solutions to the coupled Schr\"odinger and Poisson equations. These solutions can exist as a dark disk that coincides with the visible disk of the Milky Way, or else as isolated substructure.  These two possibilities have distinct observational signatures. To probe this, we will consider primarily two representative disk solutions, with parameters and properties given in table \ref{tab:table}.

  To simplify the discussion, we will fix $\hat{\psi}(0)=1$ in what follows.
  
 \begin{table*}[t!]
\begin{tabularx}{1\textwidth}{Y|Y|Y|Y|Y|Y}
   \hline\hline
System & Model parameters $\{ m [ {\rm eV} ],\lambda\}$ &  Squeezing Parameter $D_{\rm sq}$  & Core radius [kpc] & Core thickness [kpc] & Characteristic Mass Scale   \\ \hline
Dark Disk Universe &  $\{ 5 \times 10 ^{-11}, 10^{-46} \}$ 	& 	 $10^4$	& 	12.8 	& 	0.128 & Central Density: $10^{-19} {\rm GeV/cm}^3$	\\ \hline
Halo Substructure &  $\{  10 ^{-5}, 10^{-46} \}$ 	& 	 $10^{25}$	& 	34 	& 	$10^{-11}$ & Total mass: $5 \times 10^{10} M_{\odot}$	\\   \hline\hline
\end{tabularx}
\caption{Representative systems for astrophysical implications of disk solutions.}
\label{tab:table}
\end{table*}

\subsection{Dark Disk Universe}

If a component of particle dark matter has non-negligible self-interactions, and a mechanism for dissipating energy, it can collapse into a disk aligned with the visible matter \cite{Fan:2013tia, Fan:2013yva}. The properties of the disk, e.g.~the thickness, can be estimated by accounting for the particle physics processes at play (e.g.~Compton cooling), but precise estimates can only be inferred from yet to be performed N-body simulations. Our analysis provides an alternative realization of such a disk in the context of ultra-light scalar dark matter, and a prediction for the density profile. In contrast with \cite{Fan:2013tia, Fan:2013yva}, the solutions found here do not require a dissipation mechanism, but instead are self-gravitating solutions supported by the non-zero self-interaction $\lambda$, as emphasized below \eqref{eq:NormalizedEq}. Interestingly, this possibility is already constrained by data, and in particular, by recent data from the Gaia telescope \cite{Brown:2018dum}.

The Gaia data provides a loose upper bound on the density of dark matter in a thin dark disk  \cite{Schutz:2017tfp,Buch:2018qdr}. Bounds are typically formulated in terms of the surface density $\Sigma_{DD}$, related to the density inside the disk $\rho$ by $\Sigma_{DD} = h_{DD} \rho$, where $h_{DD}$ is the disk thickness. The current bound on the surface density is given by $\Sigma_{DD} \lesssim 5 M_{\odot}/{\rm pc}^2 $ \cite{Schutz:2017tfp,Buch:2018qdr}, which of course only applies provided the core radius stretches out to kpc scales, $R_c>{\rm kpc}$ in equation \eqref{Rc}.  This translates to a bound on the self-interaction and the mass, independent of $D_{\rm sq}$,
\be
\label{mlower}
\left(\frac{m}{\rm eV}\right)^4 \frac{1}{\lambda^{3/2}} \lesssim 10^{48}.
\ee
For a given mass $m$, the Gaia constraint thus provides a lower bound on $\lambda$. For example, for $m=10^{-5}$ eV and with $D_{\rm sq}$ fixed to give a disk of radius $R_c={\rm kpc}$, the Gaia constraint is a lower-bound on the self-interaction $\lambda > 10^{-46}$.

Interestingly, saturating the bound \eqref{mlower} can correspond to a disk which comprises a very small fraction of the total dark matter. For the sake of comparison, we take as a benchmark mass that which is contained within the virial radius of the Milky Way. The ratio to the disk mass is given by,
\be
f_{DD} \equiv \frac{M_{DD}}{M_{vir}}=1.1 \times 10^{-49} \, \frac{D_{\rm sq}}{\sqrt{\lambda}},
\ee
which follows from equation \eqref{MDD} and the NFW profile (truncated at $R_{s}$) $ M_{vir} \sim (4 \pi/3 ) R_s^3 \rho_0 ^2$, which for the Milky Way gives $M_{vir} \simeq 3.03 \times 10^{68} {\rm GeV}$. For example, with $\lambda = 10^{-46}$ and $D=10^4$, the disc is a negligible fraction of the total dark matter. One could instead consider $D=10^{24}$, which gives a percent level fraction of dark matter in the disk.

A more relevant comparison for dark matter direct detection is the fraction of the local dark matter density that is due to the disk. The sun's position is 26 pc above the galactic plane at a radial distance of $8$ kpc \cite{Gillessen:2008qv}, which differ by a factor of $\approx 10^2$, while the relative flattening of the disk solutions here is $\sqrt{D_{\rm sq}}$. Thus a disk which reaches our radial position and with $D \lesssim 10^4$ can in principle have a non-negligible contribution. Quantitatively, for the example of  $\lambda = 10^{-46}$, $D=10^4$, $m=5 \times 10^{-11} {\rm eV}$, the local density is $\approx 10^{-19} {\rm GeV}/{\rm cm}^3$, making the disk a negligible contribution to the total local dark matter density. For larger values of $D_{\rm sq}$, the local dark matter density is totally unaffected due to the increased compression.

This is not to say that these disks are necessarily astrophysically uninteresting. In particular, this has implications for fuzzy dark matter \cite{Hu:2000ke,Hui:2016ltb}. Namely, the bound \eqref{mlower}, which follows from explicit solutions found here, allows for astrophysically interesting dark disks, i.e.~with a thickness $>10{\rm pc}$ and $\Sigma_{DD} \gtrsim 10 M_{\odot}/{\rm pc}^2$, even for the requisite small mass range $m\sim 10^{-22}-10^{-21} {\rm eV}$ \cite{Hu:2000ke,Hui:2016ltb}.  Demanding $\Sigma_{DD} \gtrsim 10 M_{\odot}/{\rm pc}^2$ translates to the bound $m/{\rm eV } \gtrsim 10^{12} \lambda^{3/8}$. For $m=10^{-21} {\rm eV}$ this is saturated for $\lambda \simeq 6\times 10^{-89}$, which is slightly larger then the $\lambda$ considered in \cite{Desjacques:2017fmf}, corresponding to a disk of thickness $240 {\rm pc}$. Thus astrophysically interesting dark disks \emph{can} exist in the fuzzy dark matter scenario, provided that $\lambda$ has an extremely small value.

This result indicates that the observation of a dark disk would not rule out axion or superfluid dark matter \cite{Hu:2000ke,Sikivie:2009qn,Hui:2016ltb,Berezhiani:2015bqa,Ferreira:2018wup,Alexander:2018lno,Alexander:2016glq,Alexander:2018fjp} in favor of dissipative dark matter \cite{Fan:2013tia, Fan:2013yva}. Ideally one could use such an observation to make a stronger statement, such as distinguishing the dark disk of the competing scenarios. The primary distinction between the two is the predicted density profile. While the ultra-light scalar dark disk has a Gaussian density profile \eqref{eq:Fitrho}, the dark disk of dissipative dark matter (DDM) is argued to have the form \cite{Fan:2013tia, Fan:2013yva},
\be
\rho_{DDM}(x,y,z) = \rho_0 e^{- r/r_{d}}  {\rm sech}^2\left( \frac{z}{2 h_{DD}}\right)
\ee
where $r$ is the radial coordinate in the $\{ x,y\}$ plane, and $r_d$ and $h_{DD}$ are parameters. It will require very high precision measurements to distinguish these two possibilities, however, this difference is complimented by the known phenomenological differences between WIMP dark matter and ultra-light scalars, for example, in their signals at direct detection experiments.


\subsection{Halo Substructure}


 We now consider the possibility that these disk solutions could exist as isolated substructures, analogous to traditional particle dark matter sub-halos.

These are in principle subject to the constraints traditionally applied to primordial black hole and MACHO dark matter \cite{2007A, 2001PhRvL, 1985ApJ,0004637X}, which generally constrain massive compact objects to be a sub-percent fraction of the mass of the halo. For concreteness, here we will consider the effect of a disc that is 1\% the mass of the host halo. For example, for a host halo of a similar mass to the Milky Way, the characteristic size of such a disc is,
\be
R_{disc} \simeq \frac{10^{29} \lambda^{3/4}}{(m/{\rm eV})^2} {\rm pc} .
\ee
This can range in size from much less than a pc to vastly more. Thus a great diversity of disk-like subhalos is possible.

An interesting probe of such substructures is strong gravitational lensing. Here we will outline some of the key features of this signal. Assuming the orientation with respect to the line of sight is random, gravitational lensing by disk-like substructure will mimic the effect of high-ellipticity sub-halos. Depending on the orientation, a disk can mimic the effect of line-like defects (vortices) or spherical sub-halos. 

\begin{figure}[h!]
\begin{center}
\includegraphics[width=0.45 \textwidth]{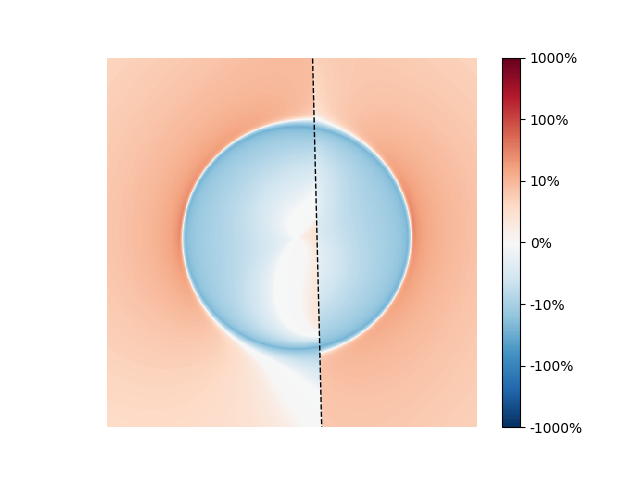}
\end{center}
\vspace{-.5cm}
\caption{Fractional change in brightness to the lensing from a spherical halo introduced by the presence of a thin disk of mass 1\% that of the halo, with position indicated by the dashed black line. The fractional change is defined as (with-without)/without.}
\label{lensing}
\end{figure}

Using the software package PyAutoLens \cite{Nightingale2015,Nightingale2018}, we simulate the lensing of a galaxy at $z = 1.0$ by a spherical halo at $z = 0.5$ and Einstein radius of 1.5 arcseconds (mass $M_{halo}\sim5\times 10^{11} M_{\odot}$), with and without a disk of mass $1\% M_{halo}$, specified by the parameters in Table \ref{tab:table}, and with orientation orthogonal to the line of sight. Figure \ref{lensing} shows the fractional change in the lensing image introduced by the presence of the disc. The change ranges from a few percent to a ten-fold increase in the brightness. While quantitative studies will need to be done to determine if this is detectable by next generation experiments such as LSST, for now this serves an indication that the morphology of substructure (such as these disk solutions) may be well probed by strong gravitational lensing.

Finally, we consider the possibility that these disk-like substructures could be the seed for satellite galaxies. Typical scales for dwarf galaxies are a radius of $\sim 100 {\rm pc}$ and a mass  $< 10^{10} M_{\odot}$~\cite{Belokurov:2006ph,SDSS2009}. For example, the dark matter dominated Hercules dwarf galaxy has a half-light radius of $\approx 330$ pc and a mass of $\approx 1.5 \times 10^7 M_{\odot}$ \cite{2011MNRAS4112118A}. While the dark disk solutions here are much flatter then the baryonic component of Hercules, a disk solution of this mass and planar extent radius does exist for $m$ and $\lambda$ satisfying $m/{\rm eV}=1.5 \times 10^{12} \lambda^{3/8}$, with a range of masses and radii available by changing $D$ and $\psi(0)$. These disk solutions could potentially seed satellite galaxies, perhaps partially, though a concrete estimate of the number density requires a detailed study of the formation process.

\vspace{-.25cm}
\section{Conclusion}
\label{sec:conclusion}

Observational evidence has demonstrated the existence of dark substructure  \cite{Hezaveh:2016ltk,Bonaca:2018fek,Price_Whelan_2018,Necib:2018iwb,Myeong:2017skt,Evans:2018bqy}. With this in mind, we have argued that disk-like solutions exist to the equations of motion describing a gravitating Bose-Einstein condensate, as emerge in models of dark matter involving ultra-light scalars and superfluids. This analysis provides a new mechanism by which observations of dark matter substructure can test the nature of dark matter.

We have not studied the evolution during or impact on structure formation, e.g.~the direct seeding of baryonic disks or the disruption of dark disk substructure by tidal forces, the latter of which may result in interesting stellar streams. These are time-dependent problems, which will be studied in future work. Another interesting avenue, and potential smoking gun substructure of superfluid dark matter, is vortices \cite{RindlerDaller:2011kx}. These will generically be present inside the dark disk solutions if the disk has a net angular momentum. We leave this, and a complete mathematical analysis of the solutions presented here, to upcoming work.

\section*{Acknowledgements}

The authors thank Michael Toomey for generating Figure~\ref{lensing}. The authors thank Robert Brandenberger, St\'{e}phane Courteau, Elisa Gouvea Ferreira, Alan Guth, David Spergel, Wayne Hu, Jeremiah P. Ostriker, and Chen Sun, for useful discussions. The authors also thank a referee for useful comments and suggestions. EM and JB are supported in part by the National Science and Engineering Research Council of Canada via PDF fellowships.

\appendix
\section{Fixed Point Algorithm}\label{app:FixedPt} 


To numerically solve the system (\ref{eq:Final}) for real-valued functions $(\psi(r),V(r))$, we implement a fixed point algorithm in MATLAB. The numerical method is as follows:
\begin{enumerate}
	\setlength\itemsep{0em}
	\item Set a sufficiently large outer radius $R$ and fix $D \geq 2$.
	\item Supply an initial guess for $\psi(r)$ on $r \in [0,R]$.
	\item Solve for $V$ in
	\[
		\bigg(\frac{\partial^2}{\partial r^2} + \frac{D}{r}\frac{\partial}{\partial r}\bigg)V = \psi^2	
	\]
	with boundary conditions $V(0) = -1$ and $V'(0) = 0$.
	\item Solve the steady-state Schr\"odinger equation
	\[
		0 = -\bigg(\frac{\partial^2}{\partial r^2} + \frac{D}{r}\frac{\partial}{\partial r}\bigg)\psi + V\psi - \psi^3 
	\]
	with Neumann boundary conditions $\psi'(0) = \psi'(R) = 0$.
	\item Iterate the previous two steps until successive solutions $\psi$ are sufficiently close in Euclidean norm. Typical convergence criteria is $10^{-5}$.
\end{enumerate}

\section{Linear Stability}\label{app:Stability} 

Let us assume that $(\psi_0(r;D),V_0(r;D))$ is a solution of (\ref{eq:Final}) for some fixed $D\geq 2$ and introduce the perturbations 
\begin{equation}\label{SolPert}
	\begin{split}
		\psi(r,t) &= \psi_0(r;D) + \varepsilon\phi(r)\mathrm{e}^{\mu t}, \\ 
		V(r,t) &= V_0(r;D) + \varepsilon w(r)\mathrm{e}^{\mu t}.
	\end{split}
\end{equation}
Here $\mu$ is the temporal eigenvalue, $\phi$ is complex-valued, and $w$ is real-valued. Putting the ansatz (\ref{SolPert}) into (\ref{eq:NormalizedEq}) and truncating at lowest order in $\varepsilon$ gives
\begin{equation}\label{Eig1}
	\begin{split}
		\mathrm{i}\mu\phi &= -\frac{\partial^2\phi}{\partial r^2} - \frac{D}{r}\frac{\partial\phi}{\partial r} + V_0\phi + \psi_0w  - \psi_0^2\bar{\phi} - 2\psi_0^2\phi  \\
		0 &= -\bigg(\frac{\partial^2}{\partial r^2} + \frac{D}{r}\frac{\partial}{\partial r}\bigg)w + \psi_0(\bar{\phi} + \phi),
	\end{split}	
\end{equation}
where $\bar{\phi}$ is the complex conjugate of $\phi$. We seek values of $\mu \in \mathbb{C}$ for which nontrivial $(\psi,w)$ can be found to satisfy (\ref{Eig1}).

System (\ref{Eig1}) is in fact an eigenvalue problem, but the term $\frac{D}{r}\frac{\partial}{\partial r}$ can be difficult to deal with numerically. Therefore, we use the identity
\bea \label{Identity}
		\bigg(\frac{\partial^2}{\partial r^2} &&+ \frac{D}{r}\frac{\partial}{\partial r}\bigg)\phi \\ 
		&&= \frac{1}{r^\frac{D}{2}}\frac{\partial^2}{\partial r^2}\bigg(r^\frac{D}{2} \phi\bigg) - \frac{1}{r^\frac{D}{2}}\bigg[\frac{D}{4}(D - 2)r^{\frac{D}{2}-2}\bigg]\phi, \nonumber
\eea
to introduce $\Phi(r) = r^\frac{D}{2}\phi(r)$ and $W(r) =  r^\frac{D}{2}w(r)$ and transform (\ref{Eig1}) to  
\bea \label{Eig2}
\mathrm{i}\mu\Phi &&= -\Phi'' + \bigg(\frac{D(D-2)}{4r^2}\bigg)\Phi + V_0\Phi + \psi_0w - \psi_0^2\bar{\Phi} - 2\psi_0^2\Phi, \nonumber  \\
		0 &&= -W'' + \bigg(\frac{D(D-2)}{4r^2}\bigg)W  + \psi_0(\bar{\Phi} + \Phi),
\eea
where $'$ denotes differentiation with respect to $r$. Separating $\Phi$ into real and imaginary parts by $\Phi = A + \mathrm{i}B$ gives
\bea
 \label{Eig3}
	\mu A &&= -B'' + \bigg(\frac{D(D-2)}{4r^2}\bigg)B + V_0B - \psi_0^2B,  \\
		\mu B &&= A'' - \bigg(\frac{D(D-2)}{4r^2}\bigg)A - V_0A - \psi_0W + 3\psi_0^2A,  \nonumber \\
		0 &&= -W'' + \bigg(\frac{D(D-2)}{4r^2}\bigg)W + 2\psi_0A.\nonumber
\eea
Now (\ref{Eig3}) is an eigenvalue problem in $(A,B)$ with the final equation acting as a constraint. 

We introduce the notation 
\[
	L := \bigg(\frac{\partial^2}{\partial r^2} - \bigg(\frac{D(D-2)}{4r^2}\bigg)\bigg),
\]
so that along with the boundary conditions $W(0) = 0$ and $W(r) \to 0$ as $r \to \infty$, we may solve for $W$ to find
\[
	W = 2L^{-1}\psi_0A.	
\] 
This reduces reduces (\ref{Eig3}) to solving 
\bea\label{EigNumerics}
		\begin{bmatrix}
			0 & -L + V_0 - \psi_0^2 \\
			L - V_0 - 2\psi_0L^{-1}\psi_0 - 3\psi_0^2 & 0
		\end{bmatrix} \cdot
		\begin{bmatrix}
			A \\ B
		\end{bmatrix} 
= \mu \begin{bmatrix}
			A \\ B
		\end{bmatrix} \nonumber
\eea
which is now a proper eigenvalue problem for $\mu$ with eigenfunction $[A,B]^T$.

Implementing the above numerically can be obtained by considering some $r_{*} \gg 1$ to restrict $r \in [0,r_{*}]$ and discretize the interval $[0,r_*]$ into equally spaced points $\{r_n\}_{n = 0}^N$ so that 
\[
0 = r_0 < r_1 < \dots < r_{N-1} < r_N = r_*,  \nonumber
\]
where $N \gg 1$ is chosen appropriately large. The boundary conditions on $W$ are implemented numerically by the Dirichlet conditions $W(0) = 0$ and $W(r_*)= 0$, and using standard finite difference approximations we have that these boundary conditions give that the numerical discretization of the linear operator $L$ is invertible. From here we again use the finite difference approximation of $L$ to write (\ref{EigNumerics}) as a matrix eigenvalue equation with $A$ and $B$ as vectors each of length $N$.

\bibliography{darkdiskrefs}

\end{document}